\documentstyle[amsfonts,aps,twocolumn,prl]{revtex}
\def\R{\mathbb{R}}
\begin{document}
\draft \title{Universal amplitudes in the FSS of three-dimensional spin models}
\author{Martin Weigel\cite{mw} and Wolfhard Janke\cite{wj}} \address{Institut
f\"ur Theoretische Physik, Universit\"at Leipzig, 04109 Leipzig, Germany, and\\
Institut f\"ur Physik, Johannes Gutenberg-Universit\"at Mainz, 55099 Mainz,
Germany} \date{\today} \maketitle
\begin{abstract}
  In a MC study using a cluster update algorithm we investigate the finite-size
  scaling (FSS) of the correlation lengths of several representatives of the
  class of three-dimensional classical O($n$) symmetric spin models on the
  geometry $T^2\times\R$. For all considered models we find strong evidence for
  a linear relation between FSS amplitudes and scaling dimensions when applying
  {\em antiperiodic} instead of periodic boundary conditions across the torus.
  The considered type of scaling relation can be proven analytically for systems
  on two-dimensional strips with {\em periodic} bc using conformal field theory.
\end{abstract}  

Conformal invariance of 2D systems at a critical point has turned out to be the
key feature for a complete, analytical description of their critical
behavior\cite{CardyBuch,HenkelBuch}.  In particular, conformal field theory
(CFT) supplies exact FSS relations {\em including the amplitudes} for these 2D
models. For strips of width $L$ with periodic boundary conditions, i.e. the
$S^1\times\R$ geometry, Cardy\cite{Cardy84a} has shown that the FSS amplitudes
of the correlation lengths $\xi_i$ of primary (conformally covariant) operators
are entirely determined by the corresponding scaling dimensions $x_i$:
\begin{equation}
  \xi_i=\frac{A}{x_i}L,
\label{amplit}
\end{equation} 
with a model independent overall amplitude $A=1/2\pi$.  This result relies on
both, the greater restrictive strength of the 2D conformal group compared with
the higher dimensional cases, which is needed for the definition of the
``primarity'' of operators, and the fact that the considered geometry is
conformally related to the corresponding flat space ${\R}^2$.

Generalizing these results to more realistic 3D geometries within the CFT
framework generically destroys the rich 2D group structure. Keeping at least the
conformal flatness condition, Cardy\cite{Cardy85} arrived at a conjecture of the
form (\ref{amplit}) for the $S^{n-1}\times{\R},\,n>2$ geometries.  Mainly for
reasons of the numerical inaccessibility of these geometries
Henkel\cite{Henkel86,Henkel87} considered the situation where even this latter
condition is cancelled: investigating the scaling behavior of the
$S=\frac{1}{2}$ Ising model on 3D columns $T^2\times{\R}$ with periodic (pbc)
or antiperiodic (apbc) boundary conditions across the torus via a transfer
matrix calculation, he found for the correlation lengths of the magnetization
and energy densities (the only primary operators in the {\em 2D} model) in the
scaling regime the ratios:
\begin{equation}
\begin{array}{rcl}
  \xi_\sigma/\xi_\epsilon & = & 3.62(7) \hspace{0.5cm} \mbox{{\em periodic}
  bc,} \\
  \xi_\sigma/\xi_\epsilon & = & 2.76(4) \hspace{0.5cm} \mbox{{\em
  antiperiodic} bc.} \\
\end{array}
\end{equation}
Comparing this to the ratio of scaling dimensions of
$x_\epsilon/x_\sigma=2.7326(16)$ a relation of the form (\ref{amplit}) seems not
to hold, {\em unless the boundary conditions are changed to be
antiperiodic}. This is in qualitative agreement with numerical work done by
Weston\cite{Weston}.

In this letter, we first revisit the Ising model on the $T^2\times{\R}$ geometry
trying to decide the exposed question with an independent Monte Carlo (MC)
method and at an increased level of accuracy. The main purpose is to investigate
further models -- in our case O($n$)$,\,n>1$ spin models --, thus adding
evidence that Henkel's result is not just a numerical ``accident'' but reflects
a universal property of such 3D systems.

\paragraph*{The model ---}
We consider an O($n$) symmetric classical spin model with nearest-neighbor,
ferromagnetic interactions in zero field with Hamiltonian
\begin{equation}
\label{Hamilton}        
{\cal H} = -J \sum_{<ij>} {\bf s}_i\cdot{\bf s}_j,\;\;{\bf s}_i \in S^{n-1}.
\end{equation}
The spins are located on a sc lattice of dimensions $(L_x,L_y,L_z)$ with
$L_x=L_y$, modeling the $T^2$ geometry by applying periodic or antiperiodic bc
along the $x$- and $y$-directions.  Effects of the finite length of the lattice
in the $z$-direction are minimized by choosing $L_z$ such that $L_z/\xi\gg 1$
and sticking the ends together via periodic bc.  As is well
known\cite{ZinnJustin}, all of these models undergo a continuous phase
transition in three dimensions, so that at the critical point the correlation
length diverges linearly with the finite length $L=L_x$.  Particular
representatives of this class are the Ising ($n=1$), the XY ($n=2$) and the
Heisenberg ($n=3$) model.

\paragraph*{The simulation ---}
For our MC simulations we used the Wolff single-cluster update
algorithm\cite{Wolff89} which is known to be more effective than the
Swendsen-Wang\cite{Swendsen} update for three-dimensional systems\cite{WJChem}.
As we want to consider antiperiodic bc for all systems in addition to the
generic periodic bc case, the algorithm had to be adapted to this situation
using the fact that in the case of nearest-neighbor interactions antiperiodic bc
are equivalent to the insertion of a seam of antiferromagnetic bonds along the
relevant boundary.

The primary observables to measure are the connected correlation functions of
the spin and the energy density:
\begin{equation}
\begin{array}{rcl}
  G_{\sigma}^c({\bf x}_1,{\bf x}_2) & = & \langle{\bf s}({\bf x}_1)\cdot{\bf
  s}({\bf x}_2)\rangle-\langle{\bf s}\rangle\langle{\bf s}\rangle, \\
  G_{\epsilon}^c({\bf x}_1,{\bf x}_2) & = & \langle\epsilon({\bf
  x}_1)\,\epsilon({\bf
  x}_2)\rangle-\langle\epsilon\rangle\langle\epsilon\rangle. \\
\end{array}
\label{conncorr}
\end{equation}
The correlation lengths $\xi_i$ in Eq.\ (\ref{amplit}) being understood as
measuring the correlations in the longitudinal $\R$-direction, one may average
over estimates $\hat{G}^c({\bf x}_1,{\bf x}_2)$ such that $({\bf x}_1-{\bf
x}_2)\parallel \hat{e}_z$ and $i\equiv|{\bf x}_1-{\bf x}_2|=\mbox{const}$, thus
ending up at estimates $\hat{G}^{c,\parallel}(i)$.  This average can be improved
by considering a zero momentum mode projection\cite{WJ93}, i.e., by correlating
layer variables made up out of the sum of variables in a given layer
$z=\mbox{const}$ instead of the original spins or local energies; this reduces
the variance by a factor of $1/L_x^2$, the influence of transversal correlations
being irrelevant for large distances $i$\cite{diplom}.

\begin{figure}[h]
  \begin{picture}(150,165)
    \put(0, 0){\includegraphics{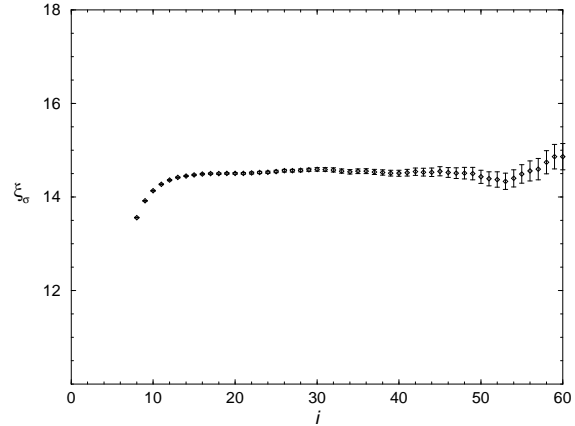}}
  \end{picture}
\caption{
Example of the set of estimators $\hat{\xi}_i$ for the magnetization density of
a $18^2\times 214$ Ising system with periodic bc.  The typical distance $\Delta$
in Eq.\ (\ref{diffmethoddelta}) was set to 8.}
\label{fig1}
\end{figure}

Assuming an exponential long-distance behavior of the correlation functions
(\ref{conncorr}), extracting the correlation lengths via a straightforward
fitting procedure requires a nonlinear three-parameter fit of the form
\begin{equation}
  G^{c,\parallel}(i)=G^{c,\parallel}(0)\exp{(-i/\xi)}+\mbox{const},
\label{corrfunction}
\end{equation}
since any numerical estimation of $G^{c,\parallel}(i)$ necessarily fails to
reproduce the correct long distance limit $G^{c,\parallel}(i)\rightarrow 0$ as
$i\rightarrow\infty$ exactly. As this amounts to an investment of the gathered
statistics into the determination of three parameters, two of which are
completely irrelevant for our ends, we used an alternative method which
intrinsically eliminates the two irrelevant parameters by using differences and
ratios of $\hat{G}^{c,\parallel}(i)$ rather than the values themselves. Given
the correlation function behaves as (\ref{corrfunction}), estimators
$\hat{\xi}_i$ for the correlation length are given by:
\begin{equation}
\hat{\xi}_i=\Delta{\left[\ln\frac{\hat{G}^{c,\parallel}(i)-\hat{G}^{c,\parallel}(i-\Delta)}
{\hat{G}^{c,\parallel}(i+\Delta)-\hat{G}^{c,\parallel}(i)}\right]}^{-1}.
\label{diffmethoddelta}
\end{equation}
The generic value for $\Delta$ is one, but it might be advantageous to choose
$\Delta>1$ in order to enhance the local drop of $G^{c,\parallel}(i)$ between
$i$ and $i+\Delta$ (the signal) against the fluctuations (the noise).  Following
this procedure one ends up with a set of estimators for the correlation length
as a function of distance $i$ as depicted in Fig.\ \ref{fig1} for the spin-spin
correlations of the Ising model: after a transition regime starting at $i=\Delta$
which is a consequence of the discreteness of the lattice as well as the above
mentioned zero momentum mode projection, the estimates settle at a plateau
indicating that the exponential long distance behavior has been reached.

The error bars in Fig.\ \ref{fig1} were generated using a combined binning and
``jackknife'' resampling scheme\cite{Efron,Berg} which is necessary due to the
strong non-linearity of the transformation (\ref{diffmethoddelta}); on the same
grounds we checked for the necessity of a bias correction. Final values for the
correlation lengths were obtained by an average of the estimators $\hat{\xi}_i$;
as neither the estimates for very small distances $i$ nor -- because of the
periodicity of the lattice in the $z$-direction -- those for distances $i\gtrsim
L_z/2$ are reliable estimates for the continuum correlation length, the range of
distances $i_{\rm min},\ldots,i_{\rm max}$ to average over was determined by a
procedure of statistical optimization, a generalized $\chi^2$-test. In order to
minimize the theoretical variance of the final average $\bar{\xi}$ each element
$\hat{\xi}_i$ was weighted by a factor proportional to a row sum of the inverse
covariance matrix, this matrix itself being again estimated by a jackknife
technique\cite{diplom,ours}.

\paragraph*{The Ising model ---}
Simulations of the Ising model were done at the most accurate estimate for the
bulk inverse critical temperature available, $\beta_c=0.2216544(3)$\cite{Talapov96},
where the influence of the given error in $\beta_c$ on the results for the
correlation lengths was checked via a temperature reweighting technique and
found negligible compared to the statistical errors; this applies to the other
models considered in this note as well.  To be able to perform a FSS analysis,
simulations were done for system sizes between $4^2\times 48$ and $30^2\times
356\approx 3\times 10^5$ sites, accumulating about eight million independent
measurements for each system.

\begin{figure}[h]
\begin{picture}(150,165)
    \put(0, 0){\includegraphics{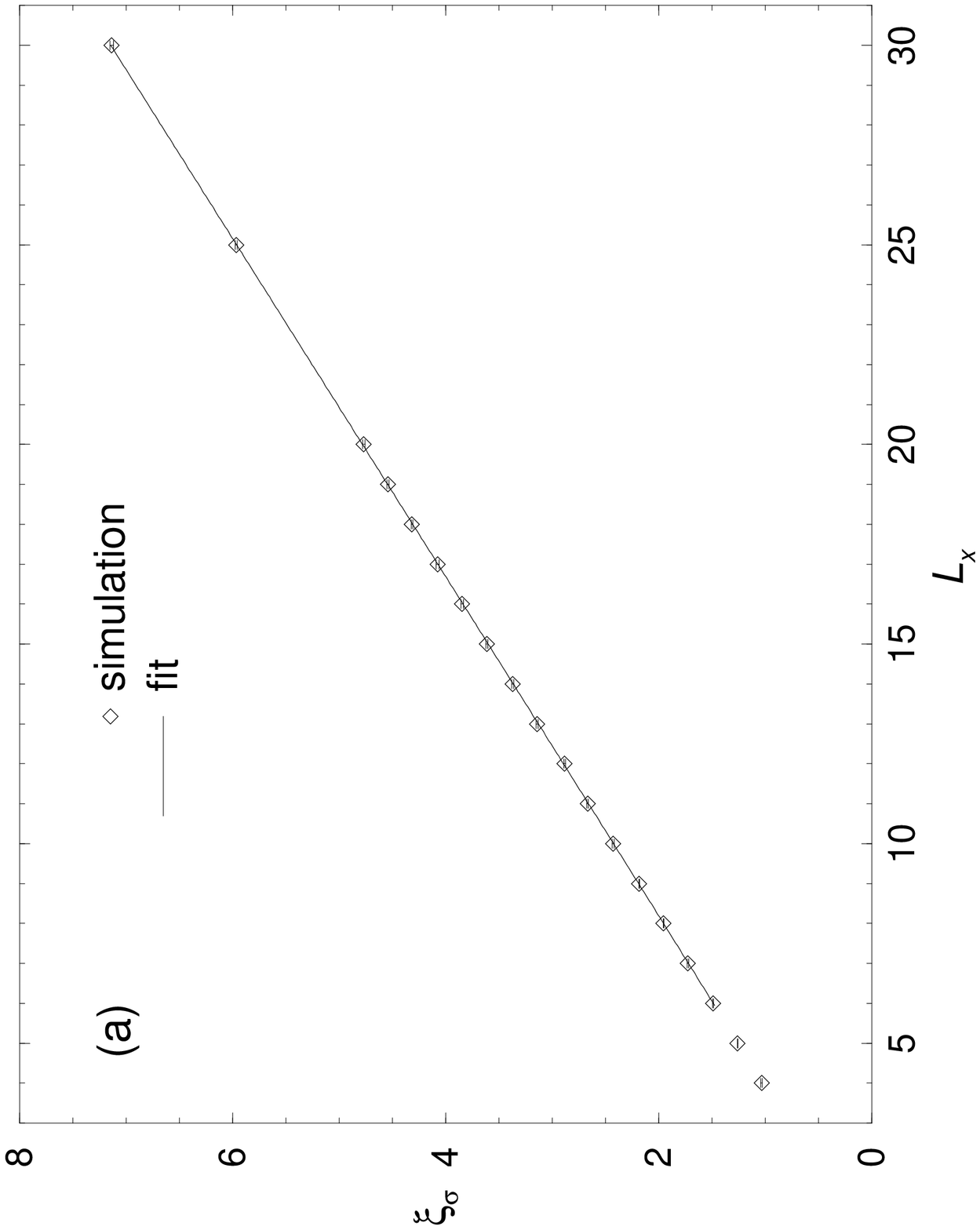}}
  \end{picture}
\begin{picture}(150,165)
    \put(0, 0){\includegraphics{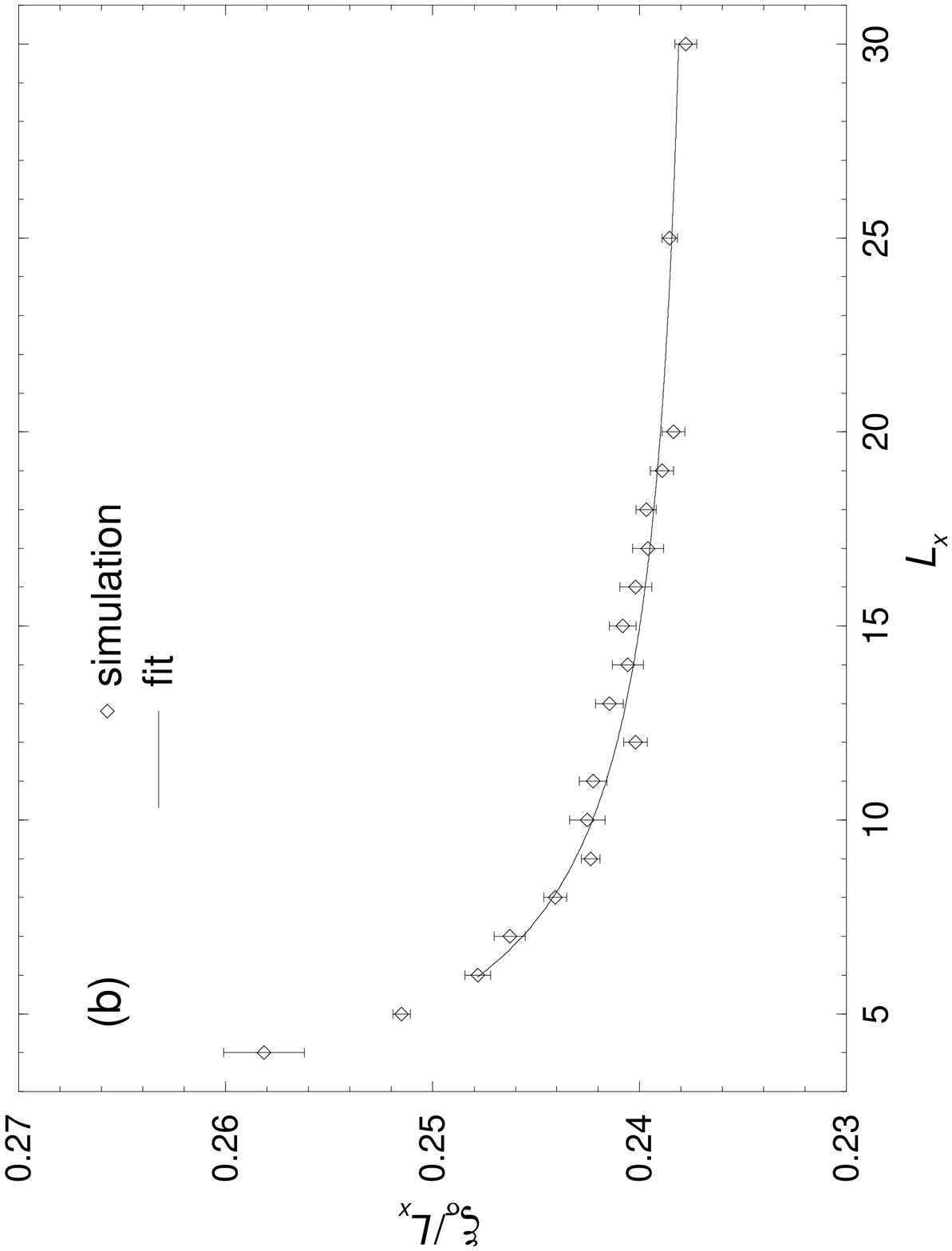}}
  \end{picture}
\caption{(a) FSS plot for the spin correlation length $\xi_\sigma(L_x)$ of the
3D Ising model with antiperiodic bc.
(b) Scaling of the amplitudes $\xi_\sigma/L_x$. Solid lines represent
least-square fits according to Eq.\ (\ref{fitform}).}
\label{fig2}
\end{figure}

As is obvious from the example in Fig.\ \ref{fig2}(a) the final estimates for
the correlation lengths show up an almost perfect linear scaling behavior as a
function of the transverse system size $L_x$. Considering the amplitudes
$\hat{\xi}/L_x$ reveals, however, that corrections to the leading linear scaling
behavior are relevant and can be clearly resolved within the accuracy of the
data, cp. Fig.\ \ref{fig2}(b).  In order to extract the leading amplitudes in
the scaling regime nonlinear fits of the form
\begin{equation}
\xi(L_x)=AL_x+BL_x^{\alpha}
\label{fitform}
\end{equation}
were done. Even though some field theoretical estimates for the correction
exponents exist\cite{ZinnJustin}, we decided to keep $\alpha$ as a parameter,
ending up at an effective correction exponent that takes higher order
corrections into account, which have some importance for the small systems;
successively dropping systems from the small $L_x$ end while monitoring the
goodness of fit parameters $\chi^2$ and $Q$ then acts as a consistency check. As
a rule, the overall corrections are negative for systems with periodic bc and
positive in the case of antiperiodic bc.

As a result of this fitting procedure we arrive at the following final estimates
for the amplitudes $A$ in Eq.\ (\ref{fitform}) and their ratios:
\begin{equation}
\begin{array}{l}
\begin{array}{l}
A_\sigma=0.8183(32) \\ A_\epsilon=0.2232(16) \\ A_\sigma/A_\epsilon=3.666(30) \\
\end{array}
\hspace{0.5cm}\mbox{for periodic bc,} \\ \\
\begin{array}{l}
A_\sigma=0.23694(80) \\ A_\epsilon=0.08661(31) \\ A_\sigma/A_\epsilon=2.736(13)
\\
\end{array}
\hspace{0.5cm}\mbox{for antiperiodic bc.}
\end{array}
\end{equation}

Comparing this to the ratio of scaling
dimensions\cite{WJChem,Bloete95,Butera97},
\begin{equation}
x_\epsilon/x_\sigma=\frac{(1-\alpha)/\nu}{\beta/\nu}=\frac{2(\nu d-1)}{\nu
d-\gamma}=2.7326(16),
\end{equation}
we find that the amplitude and exponent ratios agree very precisely in the case
of antiperiodic bc across the torus, while in the periodic case they differ by
an amount of some thirty sigma. In comparison to a first exploration by
Weston\cite{Weston}, who found ratios of about $3.7$ for periodic and $2.6$ for
antiperiodic bc, the precision could be increased by over an order of magnitude.

\paragraph*{XY and Heisenberg model ---}
Although being stringent in itself, up to this point the above result is a
singular, maybe casual, statement for the special case of the Ising
model. Believing in a universal law needs a broader backing with successful
examples, two of which being considered here.

Simulations for the XY and Heisenberg model were done at the estimated inverse
critical temperature values $\beta_c=0.4541670(32)$ and $\beta_c=0.693004(7)$,
respectively, which are weighted means of recent literature
estimates\cite{WJChem,Butera97,Ballesteros,Gottlob}. Investing about three years
of workstation time for both models altogether and using the same system sizes
as in the Ising case, we took between four and eighteen million independent
measurements for each system, both for periodic and antiperiodic bc.  Applying
the outlined tools of data analysis we arrive at scaling and amplitude plots
similar to those in Fig.\ \ref{fig2}.  Traversing the above described fitting
procedure leads to final estimates for the amplitudes $A_\sigma$ and
$A_\epsilon$ according to Eq.\ (\ref{fitform}), which are shown in Table\
\ref{tab1}.  Comparing the results for the ratios $A_\sigma/A_\epsilon$ with the
ratio $x_\epsilon/x_\sigma$ of scaling dimensions, we arrive at a highly precise
agreement for the case of antiperiodic bc and an obvious divergence in the
standard periodic bc situation for both, the XY and the Heisenberg model.  Thus
a linear relation between scaling amplitudes and scaling dimensions according to
Eq.\ (\ref{amplit}) is almost certainly valid for three generic, non-trivial
examples of 3D spin models, and one might well assume, that it is satisfied for the
whole class of O($n$) spin models, a view which is supported by further
simulations for the $n=10$ case\cite{ours} and an analytic result for the
limiting case $n\rightarrow\infty$\cite{Henkel88}, which is known to be
equivalent to the spherical model\cite{Stanley}.  In view of the analogous 2D
results it is not too far fetched, then, to argue that the numerical results
provide evidence that this relation might be of a universal, model independent
kind.

\begin{figure}[h]
  \begin{picture}(150,165)
    \put(0, 0){\includegraphics{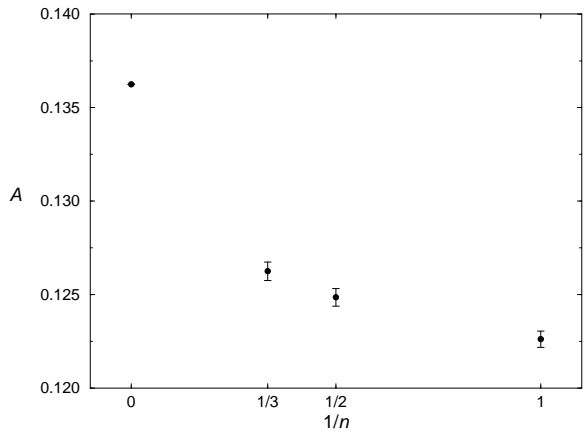}}
  \end{picture}
\caption{Amplitudes $A$ according to Eq.\ (\ref{metaamp})
versus the inverse dimension of the
order parameter $1/n$ for $n=1,2,3,\infty$.}
\label{fig3}
\end{figure}

\paragraph*{Universal amplitudes ---}
Given the scaling amplitudes for the 3D systems with antiperiodic bc behave
according to Eq.\ (\ref{amplit}), one may ask further, what the amplitude $A$ in
Eq.\ (\ref{amplit}), that was $1/2\pi$ in the 2D case, becomes in the 3D scenario
and, furthermore, if it holds true that it is universal in the sense that all
the model-dependent information is condensed in the scaling dimensions
$x_i$. The transfer matrix approach cannot give an answer to this question,
because in the Hamiltonian limit amplitudes are only given up to an overall
normalization factor.  Supposed such a relation holds, from the above results
one can give an estimate for these amplitudes using the amplitudes $A_\sigma$,
which are usually more accurate than $A_\epsilon$. Using the scaling dimensions
of the spin of $x_\sigma=0.5175(5)$, $x_\sigma=0.5178(15)$ and
$x_\sigma=0.5161(17)$ for the Ising, the XY and the Heisenberg model,
respectively, one has:

\begin{equation}
A=A_\sigma x_\sigma=\left\{
\begin{array}{l@{\hspace{0.5cm}}l}
0.12262(43) & \mbox{Ising} \\
0.12486(47) & \mbox{XY} \\
0.12625(49) & \mbox{Heisenberg} \\
\end{array}
\right. .
\label{metaamp}
\end{equation}

Taking into account the corresponding amplitude of the spherical model, which is
$A\approx 0.13624$\cite{Henkel88,Allen93}, and comparing the variation of these
values with the given errors, as is shown in Fig.\ \ref{fig3}, it becomes clear
that these amplitudes do in fact depend on the model under consideration and
seem to vary smoothly and monotonically with the dimension $n$ of the order
parameter.

To summarize, the amplitudes of the FSS of the correlation lengths of the
magnetization and energy densities of O($n$) spin models are linearly related to
the corresponding scaling dimensions for the $T^2\times\R$ geometry when
choosing {\em antiperiodic} instead of periodic bc across the torus; the amplitudes
of this relation themselves depend, in contrast to the 2D case, on the model under
consideration.

Note, however, that again in contrast to the 2D case, where the
influence of boundary conditions on the operator content has been extensively
explored\cite{Cardy86}, it is theoretically not understood up to now, why using
antiperiodic bc in 3D should restore the 2D situation.

\begin{table}[h]
\caption{FSS amplitudes of the correlation lengths of the Ising, XY, and
Heisenberg models on the $T^2\times\R$ geometry.}
\begin{tabular}{clll}
model & & \multicolumn{1}{c}{pbc} & \multicolumn{1}{c}{apbc} \\ \hline
      & $A_\sigma$    & 0.8183(32) & 0.23694(80) \\
      & $A_\epsilon$  & 0.2232(16)  & 0.08661(31) \\
\raisebox{1ex}[-1ex]{Ising} & $A_\sigma/A_\epsilon$ & 3.666(30) & 2.736(13) \\ 
      & $x_\epsilon/x_\sigma$ &  \multicolumn{2}{c}{2.7326(16)} \\ \hline
      & $A_\sigma$    & 0.75409(59) & 0.24113(57) \\
      & $A_\epsilon$  & 0.1899(15)  & 0.0823(13) \\ 
\raisebox{1ex}[-1ex]{XY} & $A_\sigma/A_\epsilon$ & 3.971(32) & 2.930(47) \\ 
      & $x_\epsilon/x_\sigma$ & \multicolumn{2}{c}{2.923(7)} \\ \hline
      & $A_\sigma$    & 0.72068(34) & 0.24462(51) \\
      & $A_\epsilon$  & 0.16966(36)  & 0.0793(20) \\
\raisebox{1ex}[-1ex]{Heisenberg} & $A_\sigma/A_\epsilon$ & 4.2478(92) & 3.085(78) \\ 
      & $x_\epsilon/x_\sigma$ &  \multicolumn{2}{c}{3.091(8)}      
\label{tab1}
\end{tabular}
\end{table}

In view of the total
lack of exact results for non-trivial 3D systems, it seems to us a rewarding
challenge for the field theorists to explain these results.

We thank K. Binder for his constant and generous support. We are grateful to
J. Cardy and M. Henkel for helpful discussions on the theoretical
background. W.J. gratefully acknowledges support from the Deutsche
Forschungsgemeinschaft through a Heisenberg Fellowship.

\end{document}